# A Hybrid Multimodal Deep Learning Framework for Intelligent Fashion Recommendation

Kamand Kalashi, Babak Teimourpour*
Department of Information Technology, Faculty of Industrial and Systems Engineering, Tarbiat Modares University (TMU),
Tehran, Iran
{ kalashi.kamand, b.teimourpour}@modares.ac.ir

*Abstract*— The rapid expansion of online fashion platforms has created an increasing demand for intelligent recommender systems capable of understanding both visual and textual cues. This paper proposes a hybrid multimodal deep learning framework for fashion recommendation that jointly addresses two key tasks: outfit compatibility prediction and complementary item retrieval. The model leverages the visual and textual encoders of the CLIP architecture to obtain joint latent representations of fashion items, which are then integrated into a unified feature vector and processed by a transformer encoder. For compatibility prediction, an "outfit token" is introduced to model the holistic relationships among items, achieving an AUC of 0.95 on the Polyvore dataset. For complementary item retrieval, a "target item token" representing the desired item description is used to retrieve compatible items, reaching an accuracy of 69.24% under the Fill-in-the-Blank (FITB) metric. The proposed approach demonstrates strong performance across both tasks, highlighting the effectiveness of multimodal learning for fashion recommendation.

## I. Introduction

The rapid expansion of online fashion retail platforms has led to an unprecedented diversity of products, making it increasingly challenging for users to identify visually coherent and stylistically compatible outfits. This complexity has underscored the growing importance of intelligent fashion recommender systems capable of understanding both the aesthetic and contextual relationships among fashion items (Han et al., 2017; Lin et al., 2019). Traditional recommender approaches, primarily designed for domains such as movies or books, often focus on individual item suggestions. However, in the context of fashion, recommendation quality depends not only on item relevance but also on outfit-level compatibility, that is, how well multiple items harmonize when combined into a single ensemble (Vasileva et al., 2018).

Two essential tasks have emerged in this domain: outfit compatibility prediction (CP) and complementary item retrieval (CIR). The former evaluates whether a set of fashion items forms a cohesive outfit, while the latter aims to retrieve missing or complementary items that complete a partially defined outfit (Han et al., 2017). Early research addressed these tasks using pairwise similarity models such as Siamese networks or conditional similarity networks (Vasileva et al., 2018). Although effective to some extent, these models primarily captured local relationships and failed to represent higher-order dependencies across multiple items. Sequential modeling approaches based on recurrent neural networks (RNNs) were later proposed (Li et al., 2020; Chen et al., 2021), but their reliance on order-sensitive architectures limited their applicability since the arrangement of fashion items within an outfit is inherently unordered.

Recent advances in transformer-based architectures and multimodal representation learning have significantly enhanced the ability to model complex visual and textual relationships in fashion data (Radford et al., 2021; Sarkar et al., 2023). In particular, the Contrastive Language–Image Pre-training (CLIP) model has demonstrated exceptional capability in aligning image and text modalities through joint contrastive learning. Leveraging CLIP's encoders allows for extracting semantically rich and visually meaningful embeddings, which can be used to model outfit compatibility and item complementarity in a unified framework.

In light of these developments, this study introduces a hybrid deep learning model that simultaneously performs outfit compatibility prediction and complementary item retrieval. The proposed framework utilizes CLIP's visual and textual encoders to obtain multimodal embeddings of fashion items, which are subsequently integrated through a transformer encoder to capture high-level inter-item relationships. For compatibility prediction, a specialized "outfit token" is introduced to represent the global context of an outfit, while in the complementary item retrieval task, a "target item token" enables the generation of latent representations for retrieving contextually relevant items.

Experimental evaluations conducted on the Polyvore dataset (Vasileva et al., 2018) demonstrate that the proposed model achieves an AUC score of 0.95 in the compatibility prediction task and 69.24% accuracy on the Fill in the Blank (FITB) metric for complementary item retrieval, outperforming several prior approaches. Overall, this study contributes a unified and scalable framework for fashion recommendation by effectively bridging the tasks of compatibility prediction and item retrieval through multimodal deep learning.

The remainder of this paper is structured as follows: Section II reviews related work, Section III outlines the methodology, Section IV presents the results and analysis, Section V concludes with key findings, limitations, and directions for future research.

## II. RELATED WORK

### A. Pairwise Item Compatibility and Type-Aware Embeddings

Early work on outfit compatibility focused on pairwise comparisons between fashion items. For example, researchers used Siamese convolutional neural networks to embed images of items from different categories into a shared "style space" in which compatible items are close while incompatible ones are distant (Veit et al., 2015). Building on this, Conditional Similarity Networks (CSNs) were introduced to model multiple aspects of similarity (e.g., category, color, style) within a unified embedding space, thereby offering more expressive compatibility modeling (Veit et al., 2017), (Saed et al., 2024).

Subsequently, the notion of type-aware embeddings was proposed (Vasileva et al., 2018) to address subtle compatibility issues such as chain interactions (e.g., shoes–hat–blouse). In this model, separate embeddings are learned for different item-type pairs (e.g., hat–blouse, hat–shoe), which helps avoid collapsing compatibility across unrelated types. More recently, SCE-Net (Tan et al., 2019) introduced a unified embedding space with hidden parallel similarity conditions, enabling dynamic weighting of different similarity dimensions and improved generalisation for new categories (Soltanshahi et al., 2025). These approaches provided a strong basis for compatibility prediction at the item-pair level; however, they are limited in modelling whole-outfit compatibility because they aggregate pairwise signals or assume fixed ordering.

### B. Global Constraints, Set-wise and Sequence-based Approaches

Recognising that outfits consist of multiple items whose mutual interactions matter, subsequent research shifted to modelling whole-set or sequence representations. For instance, Han et al. (2017) modelled an outfit as a sequence of items and used a bidirectional LSTM to learn transitions between items in the sequence. While this approach captures global constraints across items, treating an outfit as an ordered sequence introduces unrealistic position bias because the permutation of items should not alter the compatibility score.

### C. Graph-based and Transformer-based Methods

More recently, graph neural networks (GNNs) and transformer-based architectures have become popular for modelling complex relations among fashion items. For example, the Node-wise Graph Neural Network (NGNN) represents each outfit as a subgraph where each node corresponds to a category and edges represent inter-category interactions; compatibility is then scored via attention mechanisms over node representations (Cui et al., 2019), (Saed et al., 2025), (Kalashi et al., 2024), (Soltanshahi et al., 2023).

This graph paradigm is able to capture richer inter-item relationships in an outfit than sequence-based or pairwise approaches. In parallel, transformer-based methods for fashion recommendation have emerged. For instance, Chen et al. (2019) introduced POG, a personalised outfit generation model that employs a transformer encoder-decoder architecture to jointly model user preferences and outfit compatibility. These architectures make it possible to represent item interactions without assuming a fixed order and to incorporate textual metadata.

### D. Complementary Item Retrieval

While compatibility prediction remains a core task, complementary item retrieval (CIR) has received growing attention. The retrieval task requires selecting missing or complementary items for a partially defined outfit, which implies a large-scale search over item databases and efficient ranking strategies. For example, Lin et al. (2020) proposed a scalable method that scores compatibility between a "target item" and each item in an incomplete outfit and sums the scores to rank candidate items; however, it treats compatibility at a pairwise level and does not consider global outfit context. More recently, transformer-based retrieval models (Sarkar et al., 2023) model all items in an incomplete outfit as a set, learn a holistic representation, and retrieve compatible items more effectively. Overall, most CIR methods still lag behind CP (compatibility prediction) in terms of integration and end-to-end training, which presents an opportunity for unified frameworks.

### E. Limitations and Gap

In summary, existing literature reveals several gaps:

- Many pairwise models ignore inter-item interactions at the outfit level, leading to sub-optimal performance when multiple items are involved.

- Sequence-based methods impose artificial ordering on items and may not respect permutation invariance of outfit sets.

- Graph-based and transformer-based methods increasingly address these issues, but few integrate compatibility prediction and complementary item retrieval into a single unified framework.

- Retrieval tasks often rely on separate scoring or ranking modules and may not fully leverage global outfit context or multimodal embeddings jointly.

- Finally, while many methods use visual embeddings, fewer fully exploit textual descriptions or category-aware features in a unified model.

In light of these observations, the next section presents our proposed hybrid model, which addresses these gaps by jointly learning compatibility prediction and item retrieval tasks in a unified multimodal transformer architecture.

## III. METHODOLOGY

In this study, a hybrid fashion recommender model is proposed based on Outfit Compatibility Prediction (CP) and Complementary Item Retrieval (CIR). The overall methodology is divided into three major parts. The first part introduces the dataset used in the experiments, describing its structure, content, and partitioning strategy (see sub-section A). The second part presents the compatibility prediction framework, whose objective is to learn a unified latent representation of an outfit capable of modeling the complex, high-level relationships among its constituent items (see sub-section B). The third part focuses on complementary item retrieval, in which a model is designed to predict an appropriate missing item given an incomplete outfit and a description of the target item (see sub-section C).

Together, these components form an integrated framework that jointly models outfit-level compatibility and item retrieval within a shared latent embedding space. The following sections explain each component in detail, describing the dataset, model architecture, and training strategies that enable the system to capture multimodal relationships between fashion items effectively.

*A. Dataset*

The dataset used in this research was first introduced by Vasileva et al. (2018). It was constructed from the Polyvore fashion website, where users could create and share outfits consisting of multiple clothing items. Each item is multimodal, containing an image, a textual description, related tags, a popularity score, and category information.

Earlier, Han et al. (2017) released a smaller version of the dataset known as Maryland Polyvore, which suffered from several limitations, including a small size, incomplete metadata, and inconsistent test splits that affected the reliability of quantitative evaluations. To address these issues, Vasileva et al. (2018) collected a new, larger dataset that includes outfit and item identifiers, precise item types, titles, detailed textual descriptions, and item images. Outfits with only a single item or missing item-type information were removed. The resulting dataset contains 68,306 outfits and 365,054 fashion items.

A crucial consideration in constructing train-test splits for this dataset is the item overlap problem, as certain items appear in multiple outfits, potentially introducing bias during model evaluation. Therefore, two splitting strategies were proposed:

- **Non-disjoint split**:
  Outfits in the training, validation, and test sets are distinct, but individual items may appear in multiple outfits.

    o 68,306 outfits, 365,054 items
    o 53,306 outfits for training
    o 10,000 outfits for testing
    o 5,000 outfits for validation

- **Disjoint split**:
  A graph-based partitioning algorithm is used to ensure that no item appears in more than one subset (train, validation, or test). Items are represented as graph nodes, and edges indicate co-occurrence in an outfit. Highly repetitive items were removed.

- o 32,140 outfits, 175,485 items
- o 16,995 outfits for training
- o 15,145 outfits for validation and testing

In this study, all experiments were conducted on the non-disjoint version, as it has been more extensively used in prior work, allowing for a more direct comparison of results with existing studies.

Since the original dataset does not contain explicit annotations for the complementary item retrieval task, a modified version of the Polyvore dataset introduced by Lin et al. (2020) (CSA-Net) was adopted. This version provides additional structures enabling the formulation of the retrieval problem.

Each version of the dataset includes several key JSON files:

- **compatibility_<train/valid/test>.json**
  Used for the compatibility prediction task. Each entry represents an outfit instance, where the first element indicates the label (1 for compatible, 0 for incompatible) followed by the item identifiers.

- **fill_in_blank_<train/valid/test>.json**
  Used for the complementary item retrieval task. Each entry is a dictionary containing information about the question (incomplete outfit), the answer (target item identifiers), and the position of the missing item.

Item images are stored according to their identifiers and organized as lists of outfit items in each split. Each image has a fixed resolution of 300 × 300 pixels, providing a consistent visual representation for processing by computer vision models. Figure 1 presents a sample of item images from the dataset, illustrating the diversity of clothing types and visual styles included in the collection.

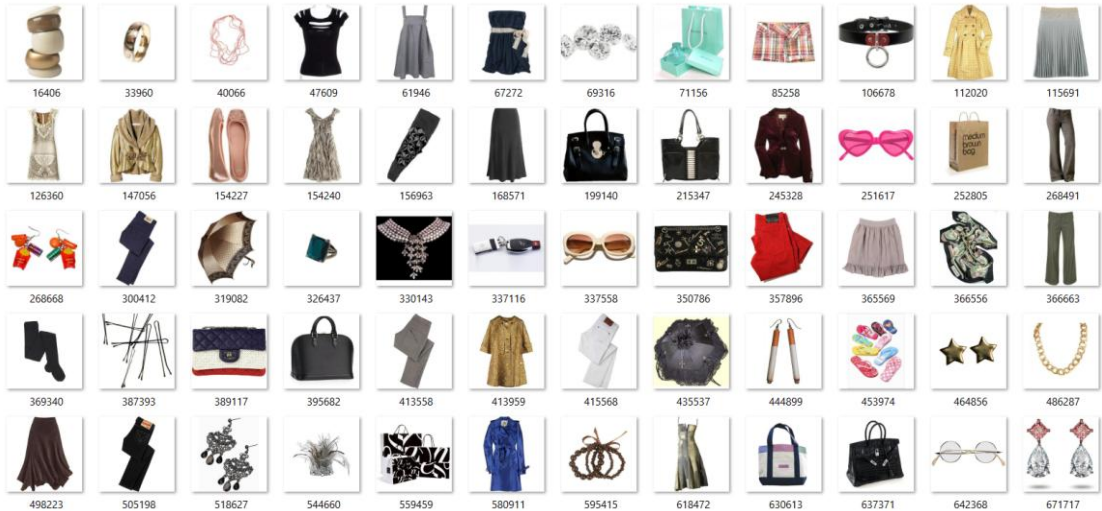

Figure 1. A sample of item images from the Polyvore dataset

B. *Outfit Compatibility Prediction*

The outfit compatibility prediction task aims to predict the degree of compatibility among all items in an outfit. Given an outfit

$$O = \{(I_i, T_i)\}_{i=1}^{L}$$

where $I_i$ denotes the image and $T_i$ the textual description of item $i$, the model learns a nonlinear function that predicts a compatibility score $c \in [0,1]$, where 1 indicates full compatibility. Item images and their textual descriptions are passed through an image encoder $E_{img}$ and a text encoder $E_{text}$ to extract visual and textual feature vectors. These are concatenated to form the item-level feature representation:

$$u_i = E_{img}(I_i) \| E_{text}(T_i),$$

where $\|$ denotes the concatenation operation. The set of feature vectors $F = \{u_i\}_{i=1}^{L}$ represents all items in an outfit.

To learn a global representation of the outfit, we introduce a learnable Outfit Token ( $x_{Outfit}$ ) that is prepended to the feature sequence $F$ and passed through a transformer encoder $E_{Trans}$. The output corresponding to this token is considered the overall outfit embedding, which is then fed into a multi-layer perceptron (MLP) to predict the final compatibility score:

$$c = MLP\big(E_{Trans}(x_{Outfit}, F)\big).$$

Unlike Vision Transformers (ViT), no positional encoding is used since the compatibility of an outfit is invariant to the order of its items. The model components $E_{Trans}$, $E_{img}$, and $E_{text}$ are trained jointly using the Focal Loss (Lin et al., 2017), which is effective for imbalanced classification problems by focusing more on hard-to-classify samples:

$$L_{Focal} = -\sum_{i=1}^{n}(1-p_i)^{\gamma} \log(p_i),$$

where $p_i$ is the predicted probability for the true class and $\gamma$ controls the focus on difficult examples.

C.    *Complementary Item Retrieval*

The complementary item retrieval task aims to find an item that both matches the existing style of an incomplete outfit and satisfies a given target description. Given an incomplete outfit and a textual description of the target item, the model generates a latent embedding for the target item that captures its compatibility with the current outfit context.

To simulate real-world retrieval scenarios where the image of the target item is not available, a Target Item Token $s$ is defined as:

$$s = x_{Img} \| E_{text}(T)$$

where $x_{Img}$ is a blank image placeholder and $E_{text}(T)$ is the textual embedding of the target description. This token is appended to the set of item feature vectors $F$ representing the incomplete outfit. The transformer encoder $E_{Trans}$ processes this input, and the output corresponding to the target token is passed through an MLP to generate the target latent vector:

$$t = MLP(E_{Trans}(s, F))$$

The model is trained using a set-based ranking loss (Lin et al., 2020), designed to ensure that the embedding of the target item is close to compatible (positive) items and far from incompatible (negative) ones. The loss function is defined as:

$$L(t, p, N) = L_{All}(t, p, N) + L_{Hard}(t, p, N)$$

with

$$L_{All}(t, p, N) = \frac{1}{|N|} \sum_{j=1}^{|N|} \left[d(t, f^p) - d(t, f_j^N) + m\right]_+$$

$$L_{Hard}(t, p, N) = \left[d(t, f^p) - \min_j d(t, f_j^N) + m\right]_+$$

where $[x]_+ = \max(0, x)$ denotes the hinge function, $t$ is the latent target vector, $f^p$ is the positive sample, $f_j^N$ represents the $j$-th negative sample, and $m$ is a predefined margin.

The loss formulation includes both all-sample and hard-sample components, allowing the model to learn discriminative features that can distinguish subtle differences between similar items (*Oh Song et al., 2016; Xuan et al., 2020*).

A pretraining stage on the compatibility prediction task is used to initialize the encoders and transformer weights, leading to significant improvements in retrieval performance. The pretrained transformer captures high-level compatibility relations, while the pretrained image encoder extracts fashion-specific visual features.

Finally, during inference, the learned encoders allow single-item embeddings to be generated, enabling efficient indexing and retrieval via K-Nearest Neighbor (KNN) search using similarity metrics such as cosine distance. This design allows large-scale retrieval across millions of fashion items efficiently.

IV. RESULTS

This section presents the experimental results of the proposed hybrid fashion recommender system. Two main tasks are evaluated: Outfit Compatibility Prediction and Complementary Item Retrieval. Both tasks were trained and validated using the non-disjoint version of the Polyvore dataset. All experiments were conducted on Google Colab Pro using an NVIDIA A100 GPU. Each model was trained for 60 epochs under identical conditions.

A. *Outfit Compatibility Prediction*

The proposed model was trained for the outfit compatibility prediction task over 60 epochs. Throughout the training process, the accuracy, Area Under the Curve (AUC), and loss metrics were monitored on both the training and validation datasets using the Weights & Biases (W&B) platform. This tool enabled real-time visualization of training progress and facilitated monitoring of the model's convergence behavior.

The recorded metrics demonstrated several consistent trends. The training loss decreased steadily as epochs progressed, while validation loss showed mild oscillations, reflecting the large scale and inherent diversity of the dataset. The AUC values showed an upward trend and gradually stabilized, indicating improved discrimination between compatible and incompatible outfits. Similarly, accuracy increased in both training and validation stages, confirming that the model effectively learned meaningful compatibility patterns among fashion items.

Table 1 compares the performance of the proposed model with several baseline approaches, including Bi-LSTM (*Han et al., 2017*), SiameseNet (*Vasileva et al., 2018*), Type-Aware (*Vasileva et al., 2018*), SCE-Net (*Tan et al., 2019*), CSA-Net (*Lin et al., 2020*), and OutfitTransformer (*Sarkar et al., 2023*). The comparison is based on the AUC metric, which measures the model's ability to distinguish compatible from incompatible outfits.

While early approaches such as Bi-LSTM modeled outfits as sequences of items, later pairwise-based methods like SiameseNet and CSA-Net relied on learning binary compatibility between item pairs and aggregating the pairwise scores to produce an overall outfit score. These models, however, are limited in capturing higher-order interactions among multiple items within an outfit.

In contrast, our transformer-based approach leverages the self-attention mechanism to learn comprehensive, higher-order compatibility relationships across all items simultaneously. The results demonstrate that the proposed model achieves an AUC of 0.95, outperforming all compared baselines. Notably, even when using only visual features, the model surpasses several prior methods that combine both image and text modalities. Incorporating textual features further enhances the performance, confirming the advantage of multimodal feature integration.

Table 1. Comparison of AUC results for outfit compatibility prediction on the Polyvore dataset

| Model | Features | AUC | Reference |
|---|---|---|---|
| BiLSTM + VSE | ResNet-18 + Text | 0.65 | (Han et al., 2017) |
| SiameseNet | ResNet-18 | 0.81 | (Vasileva et al., 2018) |
| Type-Aware | ResNet-18 + Text | 0.86 | (Vasileva et al., 2018) |
| SCE-Net | ResNet-18 + Text | 0.91 | (Tan et al., 2019) |
| CSA-Net | ResNet-18 | 0.91 | (Lin et al., 2020) |
| OutfitTransformer | ResNet-18 + Text | 0.93 | (Sarkar et al., 2023) |
| **Proposed Method** | CLIP Image Encoder + Text | **0.95** | — |

The superior performance of the proposed model confirms the capability of transformer-based architectures in learning global outfit-level relationships and demonstrates the effectiveness of integrating multimodal embeddings derived from CLIP encoders.

*B.  Complementary Item Retrieval*

For the complementary item retrieval task, the model was trained using the same general configuration as in the compatibility prediction experiment. Each training epoch monitored the accuracy and loss metrics across the training and validation datasets. Using the W&B platform, a consistent reduction in loss and a steady increase in accuracy were observed, indicating stable convergence of the retrieval model.

As in the previous task, the experiments were conducted for 60 epochs on the Polyvore dataset. Monitoring the training curves helped identify potential overfitting or underfitting behaviors. The results showed a gradual improvement in validation accuracy and a consistent decrease in loss, confirming the model's ability to effectively learn the retrieval relationships.

The complementary item retrieval task corresponds to the Fill-in-the-Blank (FITB) problem, where the objective is to select the most compatible item from a set of candidates to complete an outfit. The evaluation metric for this task is accuracy (%)

Several strategies were explored, including pretraining on the compatibility prediction task, applying curriculum learning, and experimenting with alternative loss formulations. These strategies led to noticeable performance improvements on the FITB benchmark.

The proposed method achieved an FITB accuracy of 69.24%, surpassing all previously reported methods on the same dataset. Table 2 summarizes the comparison results.

Table 2. Comparison of FITB accuracy results for complementary item retrieval on the Polyvore dataset

| Model | FITB Accuracy (%) | Reference |
|---|---|---|
| Type-Aware | 57.83 | (Vasileva et al., 2018) |
| SCE-Net Average | 59.07 | (Tan et al., 2019) |
| CSA-Net | 63.73 | (Lin et al., 2020) |
| OutfitTransformer | 67.10 | (Sarkar et al., 2023) |
| **Proposed Method** | **69.24** | Ours |

The performance improvement can be attributed to two main factors. First, the use of CLIP-based multimodal encoders enables a more discriminative joint representation space that aligns visual and textual modalities effectively. Second, the transformer encoder captures complex compatibility relationships among all items in the outfit, providing a stronger contextual understanding during retrieval.

Overall, the experimental results demonstrate that the proposed hybrid framework achieves state-of-the-art performance on both outfit compatibility prediction and complementary item retrieval, validating its effectiveness as an integrated approach to fashion recommendation.

## V.  CONCLUSIONS AND FUTURE WORK

This study introduced a hybrid deep learning framework for fashion recommendation that jointly performs outfit compatibility prediction and complementary item retrieval. Unlike prior works that addressed these tasks independently, the proposed model employs a transformer-based architecture with task-specific tokens to learn a unified representation of entire outfits. This shared embedding captures higher-order relationships among items and enables the model to reason about both visual and textual compatibility within a single framework.

By leveraging the CLIP model as a multimodal encoder, the system effectively integrates visual and semantic features, resulting in a substantial improvement in predictive accuracy. The proposed model achieved an AUC of 0.95 for compatibility prediction and an FITB accuracy of 69.24 percent for complementary item retrieval, outperforming existing methods. The use of a single latent representation for all items reduced computational complexity and indexing requirements, making the model both scalable and suitable for large fashion datasets.

Despite these promising results, several limitations remain. The computational cost of generating CLIP embeddings for hundreds of thousands of images is high, and limited access to advanced GPU resources slowed training and experimentation. Furthermore, the absence of standardized metrics for computational efficiency in related works made direct performance comparisons difficult.

Future research can extend this work in several directions. The retrieval mechanism may be enhanced to recommend multiple complementary items simultaneously, better reflecting real-world outfit creation. Additionally, optimizing embedding generation through lightweight encoders or compression techniques, and exploring incremental learning for adapting to new fashion trends without retraining, can further improve efficiency and applicability. Overall, the findings highlight the potential of multimodal, transformer-based learning for building intelligent and scalable fashion recommender systems.